\documentclass[11pt,preprint]{aastex}

\usepackage{psfig}

\def\avg#1{\langle #1 \rangle}
\def\ecr{{W}_{\rm CR}}
\def\nsn{{\cal N}_{\rm SN}}

\def\nsn{{\cal N}_{\rm SN}}
\def\decr{\dot{W}_{\rm CR}}
\def\dnsn{\dot{\cal N}_{\rm SN}}
\def\ecrsn{\epsilon}

\def\imf{\xi}
\def\msol{\hbox{$M_\odot$}}

\def\ecut{E_{\rm cut}}
\def\etal{{\it et al.~}}
\def\iso#1#2{\mbox{${}^{#2}{\rm #1}$}}
\def\be#1{\iso{Be}{#1}}
\def\li#1{\iso{Li}{#1}}
\def\b1#1{\iso{B}{1#1}}

\def\pd{\partial}
\def\ee#1#2{#1 \times 10^{#2}}

\def\pref#1{(\ref{#1})}
\def\pcite#1{\cite{#1}}

\def\beq{\begin{equation}}
\def\eeq{\end{equation}}
\def\beqar{\begin{eqnarray}}
\def\eeqar{\end{eqnarray}}

\begin{document}

\title{STANDARD COSMIC RAY ENERGETICS AND LIGHT ELEMENT PRODUCTION}
 
\author{Brian D. Fields}
\affil{Department of Astronomy, University of Illinois,
Urbana, IL 61801, USA}

\author{Keith~A.~Olive}
\affil{TH Division, CERN, Geneva, Switzerland \\
also Theoretical Physics
Institute, School of Physics and Astronomy, \\
University of Minnesota, Minneapolis, MN 55455 USA}

\author{Michel Cass\'{e}}
\affil{Service d'Astrophysique, CEA, Orme des Merisiers,
91191 Gif sur Yvette, France \\
also Institut d'Astrophysique, 98 bis Boulevard Arago,
Paris 75014, France} 

 \and \author{Elisabeth Vangioni-Flam}
\affil{Institut d'Astrophysique, 98 bis Boulevard Arago,
Paris 75014, France}

\begin{abstract}

\vskip-6.1in
\begin{flushright}
UMN-TH-1924/00 \\
TPI-MINN-00/48 \\
CERN-TH/2000-288\\
astro-ph/0010121 \\
October 2000
\end{flushright}
\vskip+5.0in

The recent observations of an approximately linear relationship
between both Be and B and
 iron in  metal poor stars has led to a reassessment of the origin of the
light elements in the early Galaxy. In addition to standard secondary
production of BeB, it is necessary to introduce a
production mechanism which is independent of the interstellar
 metallicity (primary), and in which freshly synthesized C,O and 
He are accelerated by supernova shock waves.  Primary mechanisms are
expected to be dominant at low metallicity. At metallicities higher than
O/H $\ga -1.75$, existing data might indicate that secondary production is
dominant. In this paper, we focus on
 the secondary process, related to the standard galactic cosmic
rays, and
 we examine the cosmic ray energy requirements
for both present and past epochs.
We find the
power input to maintain
the present-day Galactic cosmic ray 
flux is about 
$\ee{1.5}{41} \ {\rm erg/s} = \ee{5}{50} \ {\rm erg/century}$;
this estimate includes energy losses from 
both the escape of 
high-energy particle and ionization losses from low-energy 
particles. 
This implies that, if supernovae are the sites of
cosmic ray acceleration, the fraction of explosion energy
going to accelerated particles 
is about $\sim 30\%$, a value which we obtain consistently 
both from considering the present cosmic ray flux and confinement
and from the present \be9 and \li6 abundances.

Using the abundances of \be9 (and \li6) in metal-poor halo stars,
we extend the analysis to show the effect of the interstellar 
gas mass on the standard galactic cosmic ray
 energetic constraints on
models of Li, Be, and B evolution.
 The efficiency of the beryllium production per erg 
may be enhanced in the past
 by a factor of about 10; thus the energetic requirement by itself
cannot be used to rule out a secondary origin of light elements. 
Only a
clear and undisputable observational determination
of the O-Fe relation in the halo will discriminate between the two
processes.

\end{abstract}


\section{Introduction}

Energetic considerations have
historically provided a fundamental constraint
on and probe of the origin of cosmic rays.
For example, energetic constraints
have been used by many authors (see, e.g., 
the monograph by Brerezinski\u{\i} \etal \pcite{bbdgp}
and references therein)
to point out that supernova explosions are
a plausible cosmic ray power source,
while other stars are not.
An accounting of the energy budget have been particularly useful 
in the case of Li, Be, and B (LiBeB) production. 
Ryter, Reeves, Gradsztajn, \& Audouze \pcite{rrga} 
used this type of argument
to rule out {\it in situ} LiBeB production in young
stars by flare particles (the ``autogenetic'' hypothesis),
in favor of the present, ``galactogenetic,'' hypothesis
of LiBeB production by galactic cosmic rays in the interstellar
medium (Reeves, Fowler, \& Hoyle \cite{rfh}).

Recently, an energetics argument has been invoked to discriminate between
two possibilities, namely, a primary or a secondary origin of light
elements in the halo (Ramaty \etal \cite{rklr}). The appearance
of a linear correlation between BeB and Fe in metal poor stars  
led to the proposal
of primary mechanisms (different from
standard GCRN) for LiBeB production,
in which C, O nuclei ejected and accelerated by
supernovae fragment on H and He at rest in the interstellar medium
(e.g., Cass\'e \etal   \cite{cac}, Ramaty \etal \cite{rkl}
Vangioni-Flam et al.\ \cite{vroc};
Higdon \etal \cite{hlr};
Ramaty \& Lingenfelter \cite{rl}).

Given the approximately linear correlation between [Be/H] and [Fe/H], 
and a mean value of Be/Fe of about $10^{-6}$, one can conclude that
if 0.1 \msol\ of iron is produced per supernova, then about 2 $\times 10^{48}$
atoms of Be are produced per supernova.  Standard secondary models of Be
production yield an energy requirement of about $\sim 5\times 10^{3}$ ergs
per Be atom at $[{\rm Fe/H}] = -2.5$ or about $10^{52}$ ergs per
supernova going into the production of Be.  Of course this number is in
excess of the total energy available in the supernova explosion. This
is the basis of the energetics problem for secondary production and will
be the main focus of this paper. 

Note that the progenitor of BeB is not Fe but O
(and C). Thus the pertinent correlation is not between BeB and Fe but
between BeB and O. If [O/Fe] is constant then the linearity between
BeB and Fe translates directly into a linear relation between BeB
and O. In this case a primary mechanism is clearly required.  However,
recent observations (Israelian \etal \cite{igr}, Boesgaard \& King
\cite{bk}, Takeda \etal \cite{tak}),  indicate that [O/H] increases with
decreasing [Fe/H]. In this case the linearity between  BeB and Fe leads
to the correlation: log(Be/H) proportional to about 1.7 [O/H], and 
should be explained by a combination of primary and secondary mechanisms 
(Fields and Olive 1999a; Fields \etal 2000, hereafter FOVC). 
While the new O/Fe results are increasing in number they are still
controversial.  
An unambiguous determination of the true
O/Fe trend in Pop II is crucial, and the results of the present
study will only further underscore the need to resolve this issue.

It has been argued 
(Ramaty \etal \cite{rklr}, Ramaty, Scully, Lingenfelter, \& Kozlovsky
\cite{rslk}, Parizot \cite{par})
that energetic arguments can be used to clarify the primary vs secondary
source for BeB.
These authors have concluded that the secondary process in the early
Galaxy meet with strong energetic difficulties. However, the 
energetics computation relies on many relatively uncertain factors and
deserves a careful analysis (including the sensitivity to the injection
energy spectrum to test the sensitivity of this criterion).
Here, we will consider the energetics of {\em standard Galactic} cosmic
rays in both the early and present Galaxy.

As we will see, a careful treatment of energetics
allows us (1) to clarify and make precise the
manner in which standard cosmic rays are to be included into
chemical evolution,
and then (2) to 
examine the viability
of proposed cosmic ray origins.
We begin by establishing (in \S \ref{sect:gcr})
the present-day cosmic ray power budget,
based on the observed properties of Galactic cosmic rays
and of the Milky Way.  By explicitly identifying supernova
remnants as the power sources for the cosmic rays (in \S \ref{sect:sne}), 
we can express the present day energetics in terms of a cosmic ray
acceleration efficiency per supernova (in \S \ref{sect:cr_eff}).
We find that this efficiency is subject to
considerable uncertainties, due to uncertainties in the
parameters describing cosmic rays, the Galaxy, and supernova
blast energies.
Nevertheless, the present efficiency must serve 
as a fiducial point against
which we compare the past energetics as determined by Be 
(\S \ref{sect:be_eff}) and \li6 (\S \ref{sect:li6}). 
In calculating the energetics of \li6 and Be, we note
the importance of the larger gas mass in the early Galaxy
relative to today, which implies a larger total number of
atoms of all kinds in the interstellar medium. 
This in turn implies a larger number of
target atoms (for a given metallicity) and thus a larger
Be production rate, than one would estimate
using the present day Galactic gas mass; the upshot is that one finds
a higher Be production efficiency by a factor of 
$M_{\rm gas,init}/M_{\rm gas,today} \sim 10$ when one takes into
account this gas-depletion effect (heretofore not considered
in previous work). 
With this factor included, we find that the cosmic ray energetics 
implied by the Be data is in good agreement with
present-day values.  Indeed, we find that the energetics
constraint is completely correlated with the basic
constraint of agreement with Be-Fe data:  models which
fit the data also do not require more energy efficiency in the past.
We discuss the implications of our results in \S \ref{sect:conclude}.

Thus, the existence of an energetics problem will ultimately rest with new
observations of Be at low metallicity.  A dispersion of points above the
secondary production predictions would indicate 
the need for a primary source of Be (perhaps occurring
in spatially localized regions) 
to explain the high points.
Barring that, the only
true discriminant between primary and secondary production of Be at this
time, will be a final and consistent determination of the O/Fe ratio at
low metallicity.

\section{Standard Cosmic Rays:  Formalism and Data}
\label{sect:gcr}

The energy budget of Galactic cosmic rays
is intimately connected with
their acceleration, propagation, and losses in the
Galaxy. Thus, we will follow the sources and sinks of
the particles and the corresponding energy inputs and losses.
The sources of galactic cosmic rays
are the events which accelerate them to their high energies.
In this section, we will simply include the sources
formally, saving comment on the physical acceleration 
mechanism (namely
supernovae) until the next section.
The sinks for standard cosmic rays 
are the losses due to escape from the Galaxy
or energy deposition (ionization losses)
in the ISM.
These sources and sinks are described within a propagation
model. 
The full, most realistic 
propagation equation is diffusive, following cosmic rays
in space as well as time.  
However, for our purposes we will only
need to follow the time history of the flux averaged
over the Galaxy.  
Thus, we turn to the simplified propagation
scheme of the ``leaky box'' model.
Further refinements are in any case unwarranted by the
approximations made in the chemical evolution portion of the
analysis.

\subsection{Cosmic Ray Formalism}

To determine the propagated, interstellar particle
spectrum we must first specify a (given) source spectrum.
The source $q_E$ is 
defined so that the number of
new particles (integrated over $4\pi$ steradians)
which are injected and accelerated
to high energy in time interval
$dt$, volume $dV$, and energy (per nucleon) range $(E,E+dE)$ 
is $d{\cal N}_{\rm inj} = q_E \ dE \; dt \; dV$;
thus $q_E$ is the number of particles accelerated 
per unit time in a unit volume
and energy band.
Similarly, the propagated particles have a spectrum
$N_E$ defined by their number in
a volume $dV$ and energy band $dE$:  $d{\cal N}_{\rm prop} = N_E \ dE \; dV$.
In other words, $N_E$ is
the cosmic ray number density in energy band $(E,E+dE)$.

In the leaky box model, the source spectrum $q_E$ and
the propagated spectrum $N_E$ are related by
\beq
\label{eq:N-prop}
\frac{\pd N_E}{\pd t} = 
  q_E - \frac{\pd}{\pd E} (b N_E) - \frac{1}{\tau} N_E
\eeq
where the $b = - \pd E/\pd t$ accounts for 
ionization energy losses and is defined so that $b > 0$;
for the numerical work below we will use 
the tabulations  of Northcliffe \& Schilling \cite{ns}
and of Janni \cite{janni}.
The effective timescale for inelastic nuclear losses and
escape is
$1/\tau = 1/\tau_{\rm esc} + 1/\tau_{\rm nuc} \simeq 1/\tau_{\rm esc}$.
Equation \pref{eq:N-prop} holds for each cosmic ray species;
of these, protons and $\alpha$-particles dominate the
energy budget.

We will assume a steady state, $\pd N_E/\pd t = 0$, which
is valid as long as $q_E$ changes slowly with respect to 
the loss timescales; this is an excellent approximation for
all but the very earliest moments in the Galaxy.
In the steady state approximation, 
eq.\ \pref{eq:N-prop} is the basic relation used to
solve for the propagated spectrum given the source spectrum
and an escape time.
It will also be useful to express $N_E$ in terms of the cosmic ray
flux $\phi_E = N_E v$, which gives
\beq  
\label{eq:phi-prop}
\frac{q_E}{\bar{\rho}} 
 - \frac{\pd}{\pd E} (\omega \phi_E) - \frac{1}{\Lambda} \phi_E = 0
\eeq
where $\omega = b/\bar{\rho}v$  and
the cosmic ray mean path is 
$\Lambda = \bar{\rho} v \tau_{\rm esc}$;
$\bar{\rho}$ is the mean density encountered by the cosmic rays.\footnote{
The mean density can be determined today by comparing 
$\Lambda$ (as measured by the ratio of cosmic ray
secondaries to primaries, e.g., B/C)
and the mean lifetime of long-lived radioactive species
in the cosmic rays (such as \be{10}).
The result 
(e.g., Simpson \& Garcia-Munoz \cite{sgm}; 
Lukasiak, Ferrando, McDonald \& Webber \cite{lfmw};
Connell \cite{connell}) 
is that $\bar{\rho} \simeq 0.3 \rho_{\rm g}$,
i.e., cosmic rays encounter a mean density about 1/3 of that
of the interstellar medium in the disk ($n_{\rm g} \simeq 1 \ {\rm cm}^{-3}$).
This is usually interpreted to mean that the cosmic rays 
spend a significant part of their lives outside of the disk
and in the halo, where the gas density is much lower.}

{}From these definitions, it follows that the
cosmic ray injection power per unit volume is 
\beq
\avg{E\; q} = A \int_0^\infty dE \ E \; q_E
\label{aveeq}
\eeq
where the mass number $A$ accounts for the fact that
$E$ is energy per nucleon.
The total cosmic ray injection power is thus
\beq
\label{eq:inject}
\decr  =  \avg{E \; q} \ V 
  =  \frac{\avg{E\; q}}{\bar{\rho}} \ \bar{M}
\eeq
where 
$V$ is the volume in which the cosmic rays propagate.
$\bar{M} = \bar{\rho} V$, the gas mass contained in
$V$, and since most of the Galaxy's mass today is in the
disk, $\bar{M} \approx M_{\rm g}$.
In fact, $\decr$ is the sum of these terms for each species in the cosmic
rays.

\subsection{Cosmic Ray Data and Energy Requirements}
\label{sect:CRreqs}

We can now estimate the energy input numerically for the
present-day Galaxy, by evaluating
each of the terms on the right side of eq.\ \pref{eq:inject}.
Of these, the factors due to cosmic rays, 
$\avg{E \; q}/\bar{\rho}$, depends on the 
mean density and on the
source spectrum we adopt.
The source spectrum is itself determined from the observed
spectrum, 
which is fairly well-determined 
at high energies by the local cosmic ray spectrum, but at
low ($< 100$ MeV/n) energies requires extrapolation and thus
is model-dependent.  

We adopt a source with a momentum (per nucleon) spectrum,
$q_p \propto p^{-s} e^{-E(p)/\ecut}$. 
This has the form of a power law in $p$, characteristic of
shock acceleration,
with a slope $2 < s < 3$.
We choose a high-energy cutoff $\ecut = 10$ TeV to account for the
maximum acceleration energy of a supernova 
(e.g., Ramaty, Scully, Lingenfelter, \& Kozlovsky \cite{rslk}).
As long as $\ecut \gg 1$ GeV, the integration of the spectrum and hence
the value of $\decr$ through eqs. (\ref{aveeq}) and (\ref{eq:inject}) is
not very sensitive to the choice of $\ecut$. That is, most of the power in
the integration comes at lower energies. To convert this to a kinetic energy
spectrum we have 
$q_E = q_p \; dp/dE = q_p/v$ 
and thus
\beq
\label{eq:spectrum}
q_E = A \ v(E)^{-1} \ p(E)^{-s} \ e^{-E/\ecut} 
    = A \ (E+m) \ (E^2 + 2mE)^{-(s+1)/2} \ e^{-E/\ecut} 
\eeq
in units where $c=1$.
This spectrum scales as $E^{-(s+1)/2}$ in the non-relativistic regime,
and as $E^{-s}$ in the relativistic regime.
The constants $A$ and $s$ are set by noting that
when energy losses can be neglected ($E \gg 100$ MeV/n), 
eq.\ \pref{eq:phi-prop} gives $q_E = \phi_E \bar{\rho}/\Lambda$.
This solution, along with measurements of $\phi_E$ and $\Lambda$,
fixes $A$ and $s$. 

To calculate the integrated spectrum, we must therefore first determine
the parameters in eq.\ \pref{eq:spectrum},
$A$ and $s$, from observations of $\phi_E$ and $\Lambda$. 
We first turn to the normalization $A$.
Based on the compilation of Mori \pcite{mori},
we take
$\phi_{E_1} =  2.8 \ {\rm cm^{-2} \ s^{-1} \ GeV^{-1}}$ 
at the (arbitrary) normalization energy $E_1 = 1$ GeV.
For the escape pathlength, we adopt the
fit of Garcia-Mu\~{n}oz \etal\ \pcite{moises},
which has $\Lambda_{E_1} = 9.2 \ {\rm g \ cm^{-2}}$.
This determines the normalization constant
in eq.\ \pref{eq:spectrum}
to be $A \simeq E_1^{s} \phi_{E_1} / \Lambda_{E_1}$.

The source spectral index $s$ is determined via
the measured values of 
$\phi_E$ and $\Lambda(E)$; at high energies,
each of these is found to behave as a power law, and thus
$q_E \sim \phi_E/\Lambda(E)$ fixes $s$.
Since $\Lambda(E)$ is itself determined by measured
cosmic ray fluxes (e.g., the B/C ratio), 
the source spectral index is best derived from a consistent
analysis which yields both $s$ and $\Lambda(E)$.
Several such analyses have been performed, giving
values for $s$ ranging from $2.15$ 
(Buckley, Dwyer, M\"{u}ller, Swordy, \& Tang \cite{RICH})
to 2.23 
(Engelmann, Ferrando, Soutoul, Gorest \& Juliusson \cite{efsgj})
to 2.3 (Webber \cite{webber})
to even 2.6 (Garcia-Mu\~{n}oz, Simpson, Guzik, Wefel, \& Margolis 
\cite{moises}). 
Here we wil adopt $s=2.2$ as a fiducial value, 
bearing in mind the allowed variation is at least
$\pm 0.1$.

Uncertainties are present in both the flux and the escape pathlength.
The parameters for the flux are sensitive to the
extrapolation   from the observed, solar system intensity to the
interstellar  intensity.  This extrapolation--``demodulation''--becomes
important at $E \le 1$ GeV, and is model-dependent.  Consequently, the
combination of measurement errors and demodualtion procedure
uncertainties  leads to variation of about a factor of 5 (see, e.g., Mori
\cite{mori}) in the quoted interstellar proton flux at 1 GeV,
$\phi_{E_1}$, and uncertainties in the interstellar spectral index
of order $\pm 0.05$ units.
For example, the analysis of Webber \cite{webber}
gives 
$\phi_{E_1} =  5 \ {\rm cm^{-2} \ s^{-1} \ GeV^{-1}}$ 
and uses a confinement which scales as $\beta R^{-0.6}$
(with $\beta = v/c$ and $R$ the rigidity)
and has $\Lambda_{E_1} = 10.7 \ {\rm g \ cm^{-2}}$; for
$s=2.1$, this
gives an $A$ about 25\% larger than our adopted value.

Both the non-relativistic and relativistic limits
show that the input power integral 
$\avg{E \; q}/\bar{\rho} = \int dE \; E \ q_E$
has its maximum contribution in the $E_{\rm max} \sim m$ regime,
and taking the dominant contribution from the relativistic limit, 
we find
\beq
\avg{E \; q}/\bar{\rho} 
  \simeq Y \ \frac{(E_1/m)^{s-2}}{s-2} \ 
         \frac{ E_1^{2} \phi_{E_1} }{\Lambda_{E_1}} 
  \simeq 5 Y \frac{E_1^{2} \phi_{E_1}}{\Lambda_{E_1}}
  =  2.1 \ {\rm GeV \ g^{-1} \ s^{-1}}
\eeq
where the latter expressions use $s = 2.2$.
The factor
\beq
Y = \sum A_i y_i^{\rm CR} \approx 1 + 4 ({\rm He/p})_{\rm CR} \simeq 1.4
\eeq
includes the additional contribution due to He (with He/H $\sim 0.1$),
with the mass number $A_i$ converting from energy per nucleon
to energy,
and where $y_i^{\rm CR} = \phi_E^i/\phi_E^p$ is the abundance of
species $i$ in the cosmic rays.
We see that the input power scales as $\phi_{E_1}/\Lambda_{E_1}$,
as it should, and the {the input power density
is fixed by local measurements of the cosmic ray flux and escape pathlength}. 
We can thus analytically estimate 
$\avg{E q}/\bar{n} =
m_p \avg{E q}/\bar{\rho} 
= 2.5 \times 10^{-19} \ {\rm erg \ yr^{-1}}$.
This simple analytical estimate is to be compared
with the results using a full
numerical integration of the source spectrum for both protons and He, 
which gives
$m_p \avg{E q}/\bar{\rho} =
4.1 \times 10^{-19} \ {\rm erg \ yr^{-1}}$
in agreement with our rough analytical estimate.\footnote{
The main difference comes from the differences
in the adopted spectrum (eq.\ \ref{eq:spectrum}):
and its relativistic approximation as $A E^{-\gamma}$
used in the analytic estimate.  When the two are both normalized 
to $\phi_{E_1}$ at $E_1$, the power law
approximation
is lower for all $E > E_1$, accounting for 
most of the discrepancy.}
If $\bar{M} \simeq M_{\rm g} \sim 10^{10} \msol$, 
then this gives
\beq
\label{eq:CRval}
\decr  =  4.9 \times 10^{48} \ {\rm erg \ yr^{-1}}
  =  4.9 \times 10^{-3} \ {\rm foe \ yr^{-1}}
\eeq
or $1.5 \times 10^{41}$ erg/s. 
This is similar to other estimates; e.g., 
Drury, Markiewicz, \& Voelk \pcite{dmv},
find $\decr = 1.1 \times 10^{41}$ erg/s,
using a $p^{-2.2}$ spectrum but somewhat different gas mass
and confinement. 
We note that, for a $p^{-2.2}$ spectrum, about 20\% of $\decr$ comes from
source energies below 1 GeV;
this fraction rapidly increases for a steeper spectra,
reaching 45\% for $p^{-2.5}$.  Many of these lower-energy particles
are stopped via ionization losses to the ISM rather than by
loss from the Galaxy. 
The contributions to particles such as these
are not counted in many estimates of $\decr$ which
only include the contributions due to escape losses,
but are important and are included in our results. 

The result for $\decr$ has significant
uncertainties tracing back to the input parameters.  As noted
above, the 
uncertainties due to cosmic ray absolute flux calibrations,
spectral index determinations, and 
demodulation corrections, which easily allow for about
a factor of 4 variation
in $A$ (Mori \cite{mori}).
As a result, $\avg{E q}/\bar{n}$ can vary by a factor of
3 within the current uncertainties.
As an illustration of the possible variations, and to 
allow comparison to previous work, we turn to the spectra of
Ramaty \etal\ \cite{rslk} and Parizot \cite{par}.
The Ramaty spectrum has $s = 2.5$, $\ecut = 10 \ {\rm GeV/amu}$,
and $\Lambda(E) = 10 \ {\rm g \ cm^{-2}}$.
To compare with our results, we choose 
the same flux normalization
as our fiducial estimate, 
$\phi_{E_1} = 2.8 \ {\rm cm^{-2} \ s^{-1} \ GeV^{-1}}$.  
In this case, 
we find 
$\avg{E q}/\bar{n} =
1.7 \times 10^{-19} \ {\rm erg \ yr^{-1}}$.
This is about 40\% of our result.
The Parizot spectrum is $q_E \propto E^{-1} e^{-E/\ecut}$,
with $\ecut$ varying from 10 MeV to 500 MeV.
This is meant to describe acceleration within superbubble
shocks.
Again, for comparison we normalize to  
$\phi_{E_1} = 2.8 \ {\rm cm^{-2} \ s^{-1} \ GeV^{-1}}$.
For $\ecut = 500$ MeV/amu, we find 
$\avg{E q}/\bar{n} =
1.2 \times 10^{-19} \ {\rm erg \ yr^{-1}}$.
which is about 1/3 of our result.
On the other hand, the $\ecut = 10$ MeV/amu is 
a low-energy spectrum which does not resemble the observed GCR flux
and should not be used in this context.
In particular, fixing the normalization to the observed
flux at our $E_1 = 1$ GeV implies that the flux at, say,
$E=10$ MeV is higher by a factor of $\exp(E_1/\ecut) = e^{100}$, 
with a similarly large increase in 
$\avg{E q}/\bar{n}$.

Finally, a significant uncertainty is the present Galactic
gas mass.  We have used $M_{\rm gas} = 10^{10} \msol$.
Smaller estimates exist:  Henderson, Jackson, \& Kerr \cite{hjk}
infer a neutral atomic hydrogen
content of $M_{\rm H I} = 4.8 \times 10^{9} \msol$;
including a helium component with
mass fraction $\sim 0.25$ gives $6.4 \times 10^{9} \msol$.
The {\em total} gas mass should include 
the molecular and ionized components, which
each are likely to be at least $\sim 25\%$
of the neutral atomic component and thus lead to the estimate
we use.
The neutral atomic mass is inferred from 21-cm radio observations,
which are themselves quite accurate, but in terms
of column densities.
To arrive at a mass requires assumptions regarding the distribution
of gas in the Galaxy, which leads to uncertainties of
at least a factor of 2.
Including the gas mass uncertainty with those in the cosmic ray
inputs, the estimate of $\decr$ can easily vary by a factor
of 5.  

To summarize: we have quantified 
the power requirement $\decr$, and 
computed the present-day value give the flux at Earth.
However we have so far made no statement about the 
physical nature of
the energy source.  We now turn to this issue.

\section{Supernovae as Sources of Cosmic Rays}
\label{sect:sne}

We now invoke supernovae as the source -- the acceleration engine -- of
cosmic rays. 
This connection has both observational and theoretical support.
Observations of supernova remnants confirm that these environments
are excellent particle accelerators.
Radio synchotron emission has long been observed for supernova remnants
and establishes that nonthermal, relativistic electrons are accelerated
in these environments.  
The detection of X-ray synchotron radiation 
(Koyama \etal\ \cite{koy}),
and TeV $\gamma$-rays (Tanimori \etal\ \cite{tanimori}) 
both confirm that the electrons are accelerated up
to $\sim 100$ TeV.  
In additional, recent $\gamma$-ray observation provide the first
direct evidence of nucleon acceleration to 
$E > 300$ MeV/n in supernova remnants
(Combi, Romero, \& Benaglia \cite{crb}).  
These observations are in broad agreement with the predictions
of theoretical models of particle acceleration in supernova shocks,
and thus a coherent picture is emerging in which it seems
that supernovae are capable of accelerating particles at least
to $\sim 100$ TeV.

The net Galactic cosmic ray injection power
due to supernovae can be written as 
\beq
\label{eq:WSN}
\dot{W}_{\rm SN} = \eta_{\rm CR} \ E_{\rm SN} \ \dnsn
\eeq
where $E_{\rm SN}$ is the average mechanical energy output of
a supernova, $\dnsn$ is the Galactic supernova rate, and 
$\eta_{\rm CR} = E_{\rm CR}/E_{\rm SN}$ is the cosmic ray acceleration
efficiency--the fraction a supernova explosion's total mechanical
energy which goes into cosmic rays.

The appearance of the supernova rate
in eq.\ \pref{eq:WSN} will allows us to 
insert the cosmic ray energy budget--and thus the cosmic ray
sources and flux--within the chemical evolution formalism
in an straightforward way.
In chemical evolution, the supernova rate is
related in a simple way to the fundamental 
input, the global star formation rate $\psi$.
Specifically,
\beq
\dnsn = \int_{\rm SN} dm \ \imf(m) \psi(t - \tau_m)
   \simeq  \psi(t) \int_{\rm SN} dm \ \imf(m)
   = X_{\rm SN} \psi / \avg{m}_{\rm SN}
\eeq
where $\imf(m)$ is the initial mass function
and $\tau_m$ is the age of a star of mass $m$,
and ``SN'' denotes the mass range of supernova
progenitors, $m\ga 8-10 \msol$.
The approximation in the last two expressions holds 
if $t \gg \tau_{m_{\rm SN}} \sim 10$ Myr,
which is an excellent approximation for all but the very
earliest moments of Galactic history.
The supernova mass fraction is 
$X_{\rm SN} = \int_{\rm SN} dm \ m \ \imf(m)/\int dm \ m \ \imf(m)$
and the mean supernova progenitor mass is
$\avg{m}_{\rm SN} = \int_{\rm SN} dm \ m \ \imf(m)/\int_{\rm SN} dm \ \imf(m)$.
For a Salpeter IMF, $X_{\rm SN} = 0.14$, and 
$\avg{m}_{\rm SN} = 19 \msol$, and thus
the supernova and star formation rates
are connected by $\dnsn = 7.4 \times 10^{-3} \psi/\msol$.
Thus, if the Galactic supernova rate is
about $(3 \pm 2) \times 10^{-2}/{\rm yr}$, as
in a recent estimate 
(Dragicevich, Blair, \& Burman \cite{dbb}), then
one would infer a star formation rate
$\psi = 4 \ \msol/{\rm yr}$, in general agreement 
with other estimates (e.g., Timmes, Diehl, \& Hartmann \cite{tdh}
and refs.\ therein;
Scalo \cite{scalo} and refs.\ therein).

Thus, we have the scaling $\dot{W}_{\rm SN} \propto \psi$, 
with the explicit relation
\beq
\label{eq:SNinject}
\dot{W}_{\rm SN} 
  = \eta_{\rm CR} \ E_{\rm SN} \ X_{\rm SN} \psi / \avg{m}_{\rm SN}
\eeq
Inserting the galactic supernova rate into
(\ref{eq:WSN}) or (\ref{eq:SNinject}), we have 
\beq
\label{eq:SNval}
\dot{W}_{\rm SN} = 3 \times 10^{-2} \eta_{\rm CR} \ {\rm foe \ yr^{-1}}
\eeq
where we use $E_{\rm SN} = 10^{51} \ {\rm erg} \equiv 1 \ {\rm foe}$,
and $\dnsn = 3 \ (100 \ {\rm yr})^{-1}$;
these are good fiducial numbers but are each subject to uncertainties
of at least a factor of 2.

\section{Cosmic Ray Efficiency in the Present Epoch}
\label{sect:cr_eff}

Now we can unambiguously link
the cosmic ray and chemical evolution formalisms.
The connection is made by asserting that supernovae
are the cosmic ray accelerators.  
To implement this statement 
we join eqs \pref{eq:inject} and
\pref{eq:SNinject}:
\beq
\label{eq:link}
\dot{W}_{\rm CR} = \dot{W}_{\rm SN}
\eeq
This equation has important implications, which we now explore.	
We first 
turn to the specific numerical implementation of this statement
and estimate the present-day cosmic ray acceleration efficiency.
We then 
elaborate on how this statement clarifies
the formalism used to date in treatments of cosmic rays in
chemical evolution.

\subsection{The Present-day Cosmic Ray Efficiency}

Equations \pref{eq:CRval}, \pref{eq:SNval}, and \pref{eq:link} 
together imply that the present-day
cosmic ray efficiency is
\beq
\label{eq:CReff}
\eta_{\rm CR} = 0.16
\eeq
i.e., 16\%, if the supernova mechanical energy is $E_{\rm SN} = 1$ foe.  
We caution that this result is not known to high
accuracy, due to the uncertainties in 
the inputs, notably the supernova energy, rate,
and present gas mass
(which also plague previous estimates).
If one allows the input parameters to vary within plausible
ranges, $\eta_{\rm CR}$ can easily span more than order of magnitude,
from $< 10$\% to 100\% or more. (Of course as $\eta_{\rm CR}$ approaches
1, we have the unphysical result that from the point of view of eq.
(\ref{eq:link}), cosmic rays require more than the available mechanical
energy in the supernova.)

However, it is encouraging that our fiducial result is
reasonable and suggestive when one recalls that
the cosmic ray energy density in the ISM is in rough equipartition
with the thermal and magnetic energy.
In other words, it is well
known that the {\em propagated} particles reach an energy balance
with their environment.
Here, we see that even {\em in the acceleration process}
there is a rough equipartition of energy between the accelerated
particles and the other components.
This empirical figure is in line with
theoretical estimates of particle
acceleration efficiencies in supernova remnants.
For example, Berezhko \& Volk \pcite{bv} recently
conclude that the acceleration efficiency could reach 50 per cent. 
It is also worth noting that this result is directly
tied to the local cosmic ray
flux.  Consequently, a similar result would be obtained in
all models which fit the observed properties of the comic ray
flux and propagation.  That is, this result is model-independent
as an order of magnitude estimate, but the precise value
depends on the details of the inputs used.

\subsection{Scaling Laws for Cosmic Ray Flux}

Equation \pref{eq:link} implies that
\beq
\label{eq:q_scale}
q = \eta_{\rm CR} \ \frac{E_{\rm SN}}{\avg{E}_{\rm inj}} \
    X_{\rm SN} \  \frac{\psi}{\avg{m}_{\rm SN} V}
\eeq
where the mean injection energy 
$\avg{E}_{\rm inj} = A \int dE \, E q_E/\int dE \, q_E$ depends
on the shape of the source spectrum.
Further, since the integrated cosmic ray flux is
proportional to the integrated source strength, $\phi \propto q$, 
we have
\beq
\label{eq:phi_scale}
\phi \propto \frac{\psi}{V}
\eeq
Thus, for a fixed volume,
a higher supernova rate 
means a higher star formation rate and
more cosmic ray acceleration;
this in turn means higher flux (for a fixed confinement).
To understand the appearance of $V$ in eq.\ \pref{eq:phi_scale},
imagine dividing the leaky box volume into two
identical subregions; 
each of these will have both a star formation rate
and a volume that is half of the larger region (recall that $\psi$
represents the total star formation rate in $\msol$/unit time and is in
effect proportional to $V$).  Thus eq. (\ref{eq:phi_scale}) is
equivalent to the statement that the star formation rate per unit volume
must be fixed as is the flux of cosmic rays. Finally, if the leaky box
volume is fixed--a good approximation today, though it may not be for
early Galaxy--then we have 
$\phi \propto \psi$, as is usually assumed 
in the literature.

It is important to note that in the early Galaxy, which gave
rise to Pop II stars, 
conditions were likely different from today.
These differences can affect not only 
cosmic ray injection, but also cosmic ray propagation.
To see this, note that the cosmic ray flux
is related to the injection and confinement
roughly by $\phi \sim \Lambda q/\bar{\rho}$,
which generalizes eq.\ \pref{eq:phi_scale} 
to 
\beq
\phi \propto \Lambda \psi/\bar{\rho} V
\eeq
Thus, the flux strength depends not only on the supernova rate
and the volume $V$ of the confinement region, but also on 
the mean escape length $\Lambda/\bar{\rho}$.
This ratio (as well as the volume) could have been different in
the early Galaxy. For example, it is very likely that the early Galaxy as
a whole had a larger volume (before it settled to a disk), it seems likely
that the cosmic ray confinement volume was  also larger.  The cosmic ray
escape time  would likely have been different as well. The escape
losses depend on the interplay between the ISM density and the Galactic
magnetic field, and  as it is unclear how these would have varied, and
consequently difficult to see whether
$\Lambda/\bar{\rho}$ would have been larger or smaller. Clearly it is
difficult to ascribe a realistic uncertainty on this quantity. 
Although such a study would be very useful, and would modify the simple
scaling of cosmic ray flux with the supernova rate, it is clearly beyond
the scope of this paper and we will  retain the 
$\phi \propto \psi$ scaling. 
Another uncertainty concerns the increased gas mass
and thus enchanced total number of ISM atoms in the halo phase 
(see below, \S \ref{sect:be_eff}). 
It could well be that the stars in which LiBeB is
observed have been formed in the thick galactic disk.

Indeed this scaling of $\phi \propto \psi$ has been made by
most previous work early Galactic cosmic ray nucleosynthesis.
Its common use is evidence for its intuitive appeal.
Note that here, we have {\em derived} this scaling.
It is demanded by considerations of energetics together with
the explicit assumption of supernovae as the 
acceleration engines of Galactic cosmic rays.
Thus we see again that considerations of energetics
help to clarify issues of cosmic ray origin.

\section{Beryllium and Cosmic Ray Efficiency}
\label{sect:be_eff}

So far, we have concentrated on cosmic ray energetics today.
Now we turn to cosmic ray energetics in the past.
Our approach follows the basic argument of Ramaty \etal \cite{rklr,rslk},
which we extend by including aspects of the early Galactic
environment that were not heretofore emphasized.
(Also, our adopted source
spectrum is somewhat different, which can be important in the 
comparison of numerical results.) 

The {\em instantaneous} cosmic ray input energy 
per supernova is
\beq
\ecrsn \equiv 
\frac{\Delta \ecr}{\Delta \nsn}
   =  \frac{\decr \, \Delta t}{\dnsn \, \Delta t } 
   =  \frac{\decr}{\dnsn}
\label{eps}
\eeq
for each epoch $t$, where 
$\ecr$ and $\nsn$ are the aggregate injection
energy and supernova number
over the Galactic history.
{}From (\ref{eps}), we can relate $\epsilon$ to the cosmic ray efficiency:
$\ecrsn =
\eta_{\rm CR} E_{\rm SN}$. To see this relation more clearly, consider
models where supernovae power cosmic ray acceleration. Then the total
injection power is fixed by  the scaling $\decr \propto q V \propto
\epsilon \psi$, normalized to the present value as calculated above.
Since $\dnsn \propto \psi$ as well, eq. (\ref{eps}) follows directly.
We will assume that $\epsilon$ is constant in time, and use
the Be-O data to check this assumption. 

The observed
Be/O ratio is connected to the energetics via
$\decr/\dot{N}_{\rm Be}$, the ratio of  the Galactic cosmic-ray power
input to the Be production rate $\dot{N}_{\rm Be}$.
As is well-described in the literature,
the total Galactic Be mass changes according to 
the usual chemical evolution expression
\beq
d M_{\rm Be}/dt = m_{\rm Be} \dot{Q}_{\rm Be} - X_{\rm Be} E 
  \simeq m_{\rm Be} \dot{N}_{\rm Be}
\eeq
Here, $\dot{Q}_{\rm Be}$ is the Be source rate, i.e.,
the total number of Be atoms produced by cosmic rays per
second.  The mass fraction of Be is $X_{\rm Be}$, and
the ejection function $E = \int dm \, \psi(t-\tau_m) \phi(m)$
quantifies the rate at which gas mass is expelled from dying stars.
The last approximate equality is valid at early times (metallicities
less than solar) when Be astration can be neglected.
Thus, at early times, 
$\decr/\dot{N}({\rm Be}) \simeq \decr/\dot{Q}_{\rm Be}$.
This ratio
is just 
the inverse of the
${Q}/W$ ratio of Ramaty \etal \cite{rslk}, up to the differences in the
observed spectra.
The $\dot{N}({\rm Be})/\decr$ ratio depends on the CNO abundances in the ISM
since $\dot{N}({\rm Be}) \propto {\rm O/H}$.  

At early times, therefore, the Be production rate
is just given by the source rate
\beq
d N_{\rm Be}/dt = \dot{Q}_{\rm Be} 
  = \sum_{ij} N_i \avg{\sigma_{ij} \phi_j}
\eeq
where the sum runs over targets $i$ and CR species $j$,
and $N_i$ is the number of Galactic {\em gas atoms}
in the form of $i$:
$N_i = n_i V$.
It is important to note that the $n_i$ refer to the gas density of
in the Galaxy proper, i.e., in regions whose material
will be incorporated into star forming regions and ultimately
in Pop II stars; this density
and is not identical to the mean density $\bar{n}$ seen
by cosmic rays on their excursions in and out of
the gas-rich parts of the Galaxy.

In standard cosmic ray nucleosynthesis, the main target
is O, and the main projectiles are protons.
Furthermore, the cosmic ray species are assumed
to occur in with the same spectral shapes, only different
relative abundances, so that $\phi_E^i = y_i^{\rm CR} \phi_E^p$.
Thus we can rewrite
\beq
d N_{\rm Be}/dt = \dot{Q}_{\rm Be} 
  = \frac{\rm O}{\rm H} N_{\rm H} \Phi_p \bar{\sigma}
\eeq
where 
$\Phi_p = \int dE \phi_E$ is the total integrated flux in protons,
and 
\beq
\bar{\sigma} = 
  \sum_{ij} A_i/{\rm O} \ y_i^{\rm CR} \ 
   \avg{\sigma_{ij} \phi}/\Phi_p
\eeq
is an abundance- and spectrum-weighted cross section,
with $A_i \in {\rm CNO}$.

Thus we have
\beq
d N_{\rm Be}/dt \sim ({\rm O/H}) N_{\rm H} \sigma \phi 
\propto ({\rm O/H}) N_{\rm H} \sigma \psi 
\eeq
that is, the product of the
number of targets (i.e., $N_{\rm O} = ({\rm O/H}) N_{\rm H}$), the appropriate
spallation cross sections, and the cosmic ray flux.
The key feature here is the appearance of $N_{\rm H}$,
the total number of H-atoms in the Galactic gas,
which encodes the fact that what matters is 
the absolute number of targets and not just their relative abundance.
Note that 
$N_{\rm H} = M_{\rm gas}({\rm H})/m_p = X_{\rm H} M_{\rm gas}/m_p$,
and since $X_{\rm H}$ varies by only a few percent over the 
history of the Galaxy, then $N_{\rm H} \propto M_{\rm gas}$.
Thus we have
\beq
\dot{Q}_{\rm Be}/\dot{W} 
  \sim \frac{({\rm O/H}) N_{\rm H} \bar{\sigma} \phi}{\avg{Eq} \bar{\rho} V}
  \sim  \frac{({\rm O/H}) N_{\rm H} \bar{\sigma}}{\avg{E}_{\rm inj} \bar{\rho} V/\Lambda}
  \sim Z M_{\rm gas}
\eeq
where the second expression uses the mean injection energy
defined after eq.\ \pref{eq:q_scale} and the approximation
$q \sim \phi/\Lambda$.
The last expression uses the chemical evolution language
of metallicity $Z \propto {\rm O/H}$ and gas mass,
and makes the usual assumption that $\Lambda$ (and thus $\bar{\rho}$)
is constant.

The appearance of $N_{\rm H}$ (or equivalently,
gas mass) in $\dot{Q}/\dot{W}$ has not been heretofore emphasized, 
but plays a key role in the energetics, since the gas content 
(absolute number of atoms) of the
Galaxy {\em decreases} with time, while the metallicity {\rm increases}.
One must include {\em both} effects to fully evaluate the 
energetics.
Figure \ref{fig:Q/W} illustrates the effect of the $M_{\rm gas}$ factor;
without it (dashed curves) one simply has $\dot{N}_{\rm Be}/\dot{W}
\propto Z$, and the plot is a straight line.  However, with it, the solid
curves show that there is a peak where the product of decreasing gas
mass and increasing metallicity is maximized.\footnote{
In the instantaneous recycling approximation, we have
$M_{\rm gas} = M_{\rm gas,init} e^{-Z/y}$, with $y$ the yield.
Thus $Q/W \sim Z e^{-Z/y}$, which is maximized at $Z = y$.}
Moreover,
at low metallicity, the curve has the same linear shape,
but is increased by the factor $M_{\rm gas,init}/M_{\rm gas,now} \sim 10$.
This large factor plays a key role in the energetics.

We can thus make a rough analysis of Be energetics in the early Galaxy.
This argument was proposed and developed by
Ramaty \etal \cite{rklr,rslk}; the detailed treatment here
closely follows and extends that of Fields \& Olive \pcite{fo99}
and FOVC.
The basic idea is that the abundances of Be
and O at any epoch are a record of the 
total cosmic ray and supernova energy inputs,
respectively, for that epoch.  
Thus an analysis of the evolution of the Be/O ratio
can shed light on the evolution of the cosmic ray energy budget.

The cosmic ray input energy per supernova is thus 
calculable given (1) the ``energy per atom'' ratio,
(2) the observed Be/O ratio, and (3) the oxygen output per
supernova.  Specifically, the following 
expression relates these quantities:
\beq
\label{eq:energetics-rate}
\frac{\decr}{\dnsn} = 
   \left[ \frac{\decr}{\dot{N}({\rm Be})} \right]_{\rm th} \ 
   \left[ \frac{\dot{N}({\rm Be})}{\dot{N}({\rm O})} \right]_{\rm obs} \
   \left[ \frac{\dot{N}({\rm O})}{\dnsn} \right]_{\rm th} 
\eeq
Note that, strictly speaking,
the $\dot{N}({\rm Be})/\dot{N}({\rm O})$ term is in fact
the ratio of Be and O production {\em rates} in the theory.  However,
as long as astration effects are unimportant and 
the Be and O {\em data} are 
related by a power law so ${\rm Be} \propto {\rm O}^{\omega_{\rm BeO}}$, then 
\beq
\label{eq:obs_fac}
\dot{N}({\rm Be})/\dot{N}({\rm O}) = 
  \omega_{\rm BeO} ({\rm Be/O})_{\rm obs}
\eeq 
In this way, one can use the observed Be trends versus O (or Fe) 
to constrain the cosmic ray energy input per supernova
over all epochs for which abundances are available.
At low-metallicities, the Be-O trend has 
$\omega_{\rm BeO} \simeq 1.7 \pm 0.2$.

Let us apply this analysis to Be-O data.
As seen in Figure \ref{fig:Be/O}, the lowest metallicity point
is $[{\rm O/H}]=-1.74$, for which ${\rm Be/H} = \ee{4.4}{-14}$.
Using Figure \ref{fig:Q/W}, the energy cost is 
$\dot{N}_{\rm Be}/\dot{W} = \ee{2}{-3} \ {\rm atom/erg}$.
The data point has ${\rm Be/O} = \ee{2.8}{-9}$ and using an
oxygen yield per supernova of $\dot{M}({\rm 0})/\dnsn = 2 \msol$
corresponding to 1.5 $\times 10^{56}$ atoms of oxygen, we find that we are
required to produce $4.2 \times 10^{47}$ atoms of Be per SN. We thus find
that the cosmic ray energy input per supernova at this metallicity is
$\ecrsn = 0.36 \ {\rm foe/SN}$, 
corresponding to an efficiency of
$\eta_{\rm CR} = 0.36$ for
$E_{\rm SN} = 1$ foe. Thus we see that on the basis of Be production
alone, one requires a total energy available per SN
that is at about the level 
required to produce the observed flux of cosmic rays. In other
words, we are able to produce the desired abundance of Be
at low metallicity with an efficiency which is comparable
that needed to accelerate cosmic rays.  

The importance of the $N_{\rm H}$ factor 
in the energy per atom calculation is now clear,
as it lowers the cosmic ray energy budget to within
values that lie within 
the wide range
allowed by present-day data and theory.
Without this factor, the energy requirements inferred
from Be would increase by about an order of magnitude.
This would put the needed energy budget beyond 
the traditional estimates.  It is worth noting, however,
that even this case is 
not out of reach, if supernova blast energies 
are more typically 3 foe, and the cosmic ray efficiencies
approach 50-60\%, notwithstanding the other uncertainties discussed in 
\S 2.2.

Let us compare the above result with the previous arguments made by
Ramaty \etal\ \pcite{rslk} and Parizot \pcite{par}. 
As mentioned in the introduction, Ramaty
\etal\  base their argument on the evolution of Be with respect to Fe. 
At the low oxygen abundance of [O/H] = -1.7, we can assume a value of
[Fe/H] = $-2.5$. Recall, that the conclusion that secondary Be production
was energetically excluded was based on the Fe yield, the constancy of
Be/Fe, and the determination of the number of Be atoms per erg produced
per supernova. At this metallicity, the Be/O ratio of 2.8 $\times
10^{-9}$ corresponds to a ratio of Be/Fe of 3.2 $\times 10^{-7}$ and is
significantly below the constant {\em mean} value of $10^{-6}$. This is a
result of the fact that the data show that the Be/Fe is not a constant
with respect to Fe/H, but decreases with decreasing metallicity (though
slowly) and that there is considerable dispersion in the data. 
Had we used the estimate of energy requirement for secondary production
of Ramaty \etal \cite{rslk} (4 $\times 10^{-4}$ atoms/erg at this metallicity),
we would have concluded that 
$(1.7 \times \ee{2.8}{-9})(\ee{1.5}{56})/\ee{4}{-4}=1.7 \times 10^{51}$ 
ergs/SN is required. 
This may or may
not be too much depending on the actual energy available.  We have
typically assumed that the total energy available in the explosion is
$10^{51}$ erg, but this is uncertain by a factor of at least 3, and if
the available energy is actually 3 times higher, there is no energetics
problem for the lowest metallicity points (and lowest Be) points observed.
It is however very likely that primary sources are required at these low
metallicities in order to explain the scatter in the data. Note that as
argued above, the energy required per Be atom is significantly reduced
when density effects are taken into account.  In that case, energetics
is unable to provide a strong indication
as to the primary vs secondary nature of Be production.

Parizot \cite{par}
also provided an energetics argument against secondary production.
We find several differences in the simple estimate which is used to
preclude secondary production.  First, at low metallicity, Parizot
assumes a Be/O ratio
of $4 \times 10^{-9}$, where it is in fact $2.8 \times 10^{-9}$
(this stems in part from his incorrect value of the solar oxygen
abundance). He further assumes that the maximum energy available for Be
production is $10^{50}$ erg/SN corresponding to an efficiency of 10\%.
This could be 3--5 times higher and as noted above, the total SN energy
could be 3 times higher.  Finally his choice of a spectrum is far  more
pessimistic than even that used in Ramaty etal 
and assumes a high energy cut off of 0.5 GeV. 
Thus from the
argument given in Parizot (2000), one can not place a
constraint on the secondary nature of Be.

For late times and high metallicities, 
the approximations of no Be astration and 
a constant Be-O slope break down.  
A more refined analysis is needed.
Given the data, what is actually observable is 
the Be/O ratio, which is in general not set by the ratio of
mass change rates
$\dot{N}({\rm Be})/\dot{N}({\rm O})$, but rather 
Be/O.
Since these observables connect not to rates but to
integral quantities, they do not allow for direct
measure of the instantaneous cosmic ray energy input
$\decr$.
Instead, the data can determine the {\em time-averaged} value
\beq
\avg{\ecrsn} 
  = \frac{\ecr(t)}{\nsn(t)}
  = \frac{\int_0^t \ dt^\prime \ \decr}{\int_0^t \ dt^\prime \ \dnsn}
  = \frac{\int_0^t \ dt^\prime \ \dnsn \ \ecrsn}{\int_0^t \ dt^\prime \ \dnsn}
\eeq
Within the model, we assume (and then test)
a constant energy budget $\ecrsn = \ecrsn_0 = const$, and thus
$\avg{\ecrsn} = \ecrsn_0$ as well.
However, 
this quantity can also be calculated in a way
that makes explicit use of the observed
Be abundances.  

We can calculate the (time-averaged) energetics as follows:
\beq
\label{eq:energetics-int}
\frac{\ecr}{\nsn} = 
   \left[ \frac{\ecr}{N({\rm Be})} \right]_{\rm th} \ 
   \left[ \frac{N({\rm Be})}{N({\rm O})} \right]_{\rm obs} \
   \left[ \frac{N({\rm O})}{\nsn} \right]_{\rm th} 
\eeq
The cosmic ray input energy per supernova is thus 
calculable given (1) this ``energy per atom'' ratio,
(2) the observed Be/O ratio, and (3)  the oxygen output per
supernova. 
In this way, one can use the observed Be trends versus O (or Fe) 
to constrain the cosmic ray energy input per supernova
over all epochs for which abundances are available.
To do this, we rewrite eq.\ \pref{eq:energetics-int},
noting that in the code all rates in eq.\ (\ref{eq:energetics-int})
are computed explicitly.  Thus we are free to rearrange terms
and write 
\beqar
\frac{\ecr}{\nsn} 
  & = & 
   \left[ \frac{N({\rm Be})}{N({\rm O})} \right]_{\rm obs} \
   \left[ \frac{N({\rm O})}{N({\rm Be})} \right]_{\rm th} \
   \ecrsn_{\rm th} \\
  & = &  
   \frac{({\rm Be/O})_{\rm obs}}{({\rm Be/O})_{\rm th}} \ \ecrsn_{\rm th} 
\label{eq:nrg-scale}
\eeqar
where $\epsilon_{\rm th}$ is the theoretical efficiency (and must be less
than 30 --50\%). Thus the true needed efficiency $\epsilon$ will exceed the
theoretical efficiency only if the observed Be/O ratios are greater than the
theoretically predicted ones.  That is if there is significant (upwards)
scatter in the Be vs O data, or there is a true excess in Be/H at low
metallicity, will we find that energetics preclude the secondary production
of Be as was argued in FOVC. 

Thus the energetics entirely
reflects the agreement of the abundance trends.  If the data are fit
within the errors, 
then the right side of eq.\ (\ref{eq:nrg-scale}) is constant in metallicity,  
consistent with the assumption of constant $\ecr/\nsn$.  
The is the case for primary models of, e.g., 
Vangioni-Flam \etal \cite{vcfo}
or 
Ramaty \etal \cite{rslk}.  
This agreement with the data {\em also} holds
case for standard GCR models with changing O/Fe, 
at least down to metallicities ${\rm [O/H]}\simeq -1.5$.
At lower metallicities, the data is not conclusive but 
seems to suggest
the need for a primary component (FOVC).
If this is borne out by improved
data, then the need for a primary component of Be
will also be reflected in a energetics problem for the standard GCR model
at these metallicities.  Careful observations at these metallicities
are thus crucial and encouraged.  In any
case, the standard GCR model certainly provides
an acceptable fit to the Be-O data above about
${\rm [O/H]}\simeq -1.5$ (or ${\rm [Fe/H]} \simeq -2.2$), and thus 
suffers no energetics problem in this regime.

To illustrate this conclusion, we can calculate the energetics
for all of the data points illustrated in Figure \ref{fig:Be/O}.
We implement eq.\ \pref{eq:nrg-scale} for each data point, using
our model to calculate $[\ecrsn / (Be/O)]_{\rm th}$, and 
the data for $(Be/O)_{\rm obs}$.  The result appears in
Figure \ref{fig:energetics}, both with and without
normalizing the Be solar abundance to its observed value.  We see that 
in the standard case, there is scatter in the data but the
mean remains close to the solar value, which is around 0.08 foe/SN.

Combining these factors, the solar beryllium and oxygen
abundances give a present-day 
cosmic ray energy budget per supernova 
which spans a range of 
\beq
\label{eq:Be_eff}
\ecrsn = 0.5 \pm 0.4 \ \mbox{foe \ \ (per SN)}
\eeq
(using the weighted mean and the $2-\sigma$ weighted
sample variance)
or an efficiency of 
$\eta_{\rm CR}=(50 \pm 40)\%$ 
for a supernova blast wave energy
of 1 foe.
This number is consistent, within uncertainties, with our estimate 
of $\eta_{\rm CR} = 16\%$ 
using the observed cosmic ray flux (eq.\ \ref{eq:CReff}).
This concordance means that the present-day Be
abundance is consistent with the energy that has gone into the cosmic rays
up to the present epoch. Thus, standard Galactic cosmic rays are able 
to account for both the present-day Be abundance as well as
the associated energy, within the described uncertainties. 

An important but somewhat technical detail concerns the normalization
of the theory curves.  The approach we have taken in this
and previous work is to
normalize the curves by demanding that the Be abundance
at solar metallicity is equal to the solar Be abundance.
Physically, this amounts to determining the scaling between
the global average Galactic flux at the present
epoch and the present flux at the solar system.
The normalization factor is about 0.5, i.e., the solar flux is slightly high.
This factor has been included in Figure \ref{fig:Q/W}.
If we did not include this factor, we would both raise
the entire Be-O curve and increase
the energy requirement by the same factor of 1/0.5 = 2.
The overall mean is thus higher in the non-normalized case,
at $1.0 \pm 0.8$ foe/SN.  This is just at the
fiducial (but poorly determined) supernova mechanical energy budget,
but of course the lack of normalization also leads to
a poorer fit to the Be-O data.  Thus we see an illustration
of the connection between energetics and the Be-O fit:
a poorer agreement with Be-O data goes hand in hand with
poorer agreement with energetics requirements.

Finally, we recall that the
energy budget we have
determined depends sensitively
on the cosmic ray spectrum adopted.
We have used a $p^{-2.2}$ source, which we consider to be a fair
representation of the experimental determinations;
however, the present cosmic ray data allow for other spectral
forms and different energetic requirements.  For example,
used by Ramaty \etal\ \cite{rslk} is more energetically
efficient by a factor of 2.5
(i.e., $\avg{Eq}$ is 40\% of the value using our adopted
spectrum), as noted
in \S \ref{sect:CRreqs}.  Thus, using that spectrum, the 
energy requirements would drop by a factor of 2.5,
both for the present cosmic rays and for the past as derived form Be.
Thus, while the {\em ratios} of past/present values would be similar, but
the absolute energy requirements would be reduced.
Using the Parizot spectrum (with 500 MeV cutoff) would
reduce this requirement yet further.

To summarize, {\em for models with supernova-powered cosmic rays,
the fit of the energetics is completely determined by,
and follows from, the fit of the 
observed Be-O data.}  Models which fit the data 
(and have $\psi \propto \dnsn$)
do not have an energetics problem in the past if they
do not have one now. 
Conversely, models which do not fit the data (in particular,
those which underpredict the Be abundance) can also
be expected to have energetic difficulties over the Be-O
regimes which are poorly fit.

\section{Lithium-6 Constraints on Energetics}
\label{sect:li6}

We can repeat this analysis for other cosmic-ray produced elements.
For \li7 and \b11, the results are complicated by the fact that
these elements are produced in other sources.
However, \li6, like Be, is a pure cosmic-ray element, and thus the
energetics evaluation should proceed as it does for Be.  
We first consider the present-day energetics (the analog of
eq.\ \ref{eq:Be_eff}).  For \li6, we have 
$\dot{Q}_{\li6}/\decr = 0.05 \ {\rm atom/erg}$
using the solar value from Figure \ref{fig:Q/W},
and $(\li6/{\rm O})_\odot = \ee{1.8}{-7}$.  
Finally, we again take an average of $2\msol$ of 
oxygen ejected
per supernova, or $\ee{1.5}{56}$ atoms.  Together, this gives
\beq
\label{eq:eff0}
\ecrsn = 0.54 \ \mbox{foe \ \ (per SN)}
\eeq
This is comparable to, albeit somewhat higher,
than the other estimates we have made.
Also, as with the Be analysis, using a source spectrum
such as that of Ramaty \etal\ \cite{rslk} 
would lower this requirement by a factor of 2.5.

Regarding the energetics of \li6 in the past,
the analysis of the previous section carries over:
the efficiency remains constant if the theory and data 
agree.  The data for \li6 in Pop II stars are difficult to
obtain, and thus this is a less definitive test.
However, as shown by Fields \& Olive \pcite{fo99b}, 
the predictions of standard GCR agree with the presently available
data.  Thus we see that the \li6 constraints on standard GCR
are quantitatively similar to those of Be, and 
demonstrate the basic consistency of the
analysis.

Note that \li6 can be made via the $\alpha + \alpha$ fusion reaction. This
production mode is always primary since most helium in the ISM
is primordial and hence predates the Galaxy.
\li6 is thus primary in both
the standard GCR scenario of particle acceleration from the ISM,
as well as in the case GCR origin
is dominated by superbubbles (Ramaty \etal \cite{rslk}).
An approach which synthesizes aspects of these two outlooks
adds to the standard GCR scenario a component
of energetic particles (EPs)
which could also
originate from superbubbles (for a review 
see Vangioni-Flam, Casse \& Audouze \cite{vca}).
The EP component 
is highly enriched in $\alpha$ particles since its
origin is in the massive star ejecta which fill the superbubbles.
Due to this enhancement, the EP component can explain
the \li6 data with an even lower energy efficiency
than that required above for GCR particles.
Further \li6 data would help distinguish between
the EP and GCR scenarios for the $\alpha$-component of
early Galactic cosmic rays.

\section{Conclusions}
\label{sect:conclude}

In this paper, we have
reviewed the past and present
energy budget for standard Galactic cosmic rays
We determine present-day cosmic ray power requirements
from the observed cosmic ray flux and confinement.
Comparing this to the fiducial $10^{51}$ erg of kinetic energy
in a supernova blast, we find the efficiency of cosmic ray 
acceleration is about
30\%. This value cannot be calculated to a precision
better than a factor of about 3, given
the uncertainties in the cosmic ray and Galactic input parameters.
Nevertheless, we find that this value is consistent with
estimates derived from present (solar) \be9 and \li6 values.

We have extended our analysis to the study the
cosmic ray acceleration efficiency in the early Galaxy, 
as encoded in the Be abundances of metal-poor stars.
We find that energy considerations are indeed powerful,
and allow us to {\em derive}
the usual $\phi \propto \psi$ scaling.
We also find that in general, the requirement that models
satisfy energetic requirements
is equivalent to the requirement that the models
produce LiBeB evolution which fits the abundances data,
if SN are cosmic ray acceleration engines.

Furthermore, constructing a careful and consistent model for the
energetics of cosmic ray acceleration and LiBeB production has an
additional benefit: it brings to the surface key underlying physical
issues and assumptions regarding early Galactic LiBeB production.
Specifically, we have emphasized that the ISM gas content, and thus
the total number of ISM atoms, was higher in the early Galaxy.  Thus,
for a fixed metallicity, the total Galactic Be production is enhanced
due to the increased total number of targets.  This effect has been
neglected up to now but can play a crucial role in the energetics: due
to this effect, the number of Be atoms produced per cosmic ray erg
should be multiplied by about a factor 10 in the early Galaxy compared
to the estimates lacking this effect.  Consequently, the energy test
is less stringent than previously thought.  Note that one could invoke
still more complications in the form of early Galactic evolution of
cosmic ray confinement properties, or a lack of equilibrium between
cosmic ray sources and sinks; further study of these issues is best
done in a consistent model of cosmic ray acceleration and propagation
in a Galaxy which evolves in time and space.  In the absence of such a
model, the current ``leaky box'' picture that has been adopted for Be
studies, with constant cosmic ray confinement parameters, leads to the
factor of $\sim 10$ increase in Be production efficiency we have
described.  With this effect, the standard GCR model
is able both to fit the Be-Fe data, and to maintain a reasonable
(and roughly constant) acceleration efficiency over the 
history of the Galaxy 
at least down to [Fe/H] $\simeq - 2.5$.

Thus we find that energetic arguments are extremely useful tools
in analyzing LiBeB production in the early Galaxy, and thereby to
examine questions of cosmic ray origin.  However, contrary to
some analyses, we find that the standard GCR scenario does
not suffer energetics problems, and thus that energetics
alone cannot rule this model out.  
Therefore, the key of the origin of early Galactic cosmic rays
and LiBeB (i.e., a primary or
secondary process?) remains in the hands of the observers.  
Specifically, it is crucial to reconcile the different measurements
of oxygen (and thus BeB-O and O-Fe) in halo stars.
The determination of reliable oxygen abundances 
is to be ranked as a first priority of the abundance observation programs.

Indeed, looking towards the future, we can hope to turn the problem around.
With reliable and accurate oxygen data, we can
establish the origin of LiBeB  and cosmic rays in the early Galaxy.
With this knowledge in hand, we can then use the detailed LiBeB abundance
patterns and scatter to learn more about the physical conditions of 
gas and accelerated particles in the early Galaxy.

\acknowledgments
 We thank
Tom Jones and Vasiliki Pavlidou for fruitful discussions.
This work
was supported in part by DoE grant DE-FG02-94ER-40823 at the University of
Minnesota.

\newpage

\centerline{\bf FIGURES}

\begin{enumerate}

\item
\label{fig:Q/W}
\psfig{file=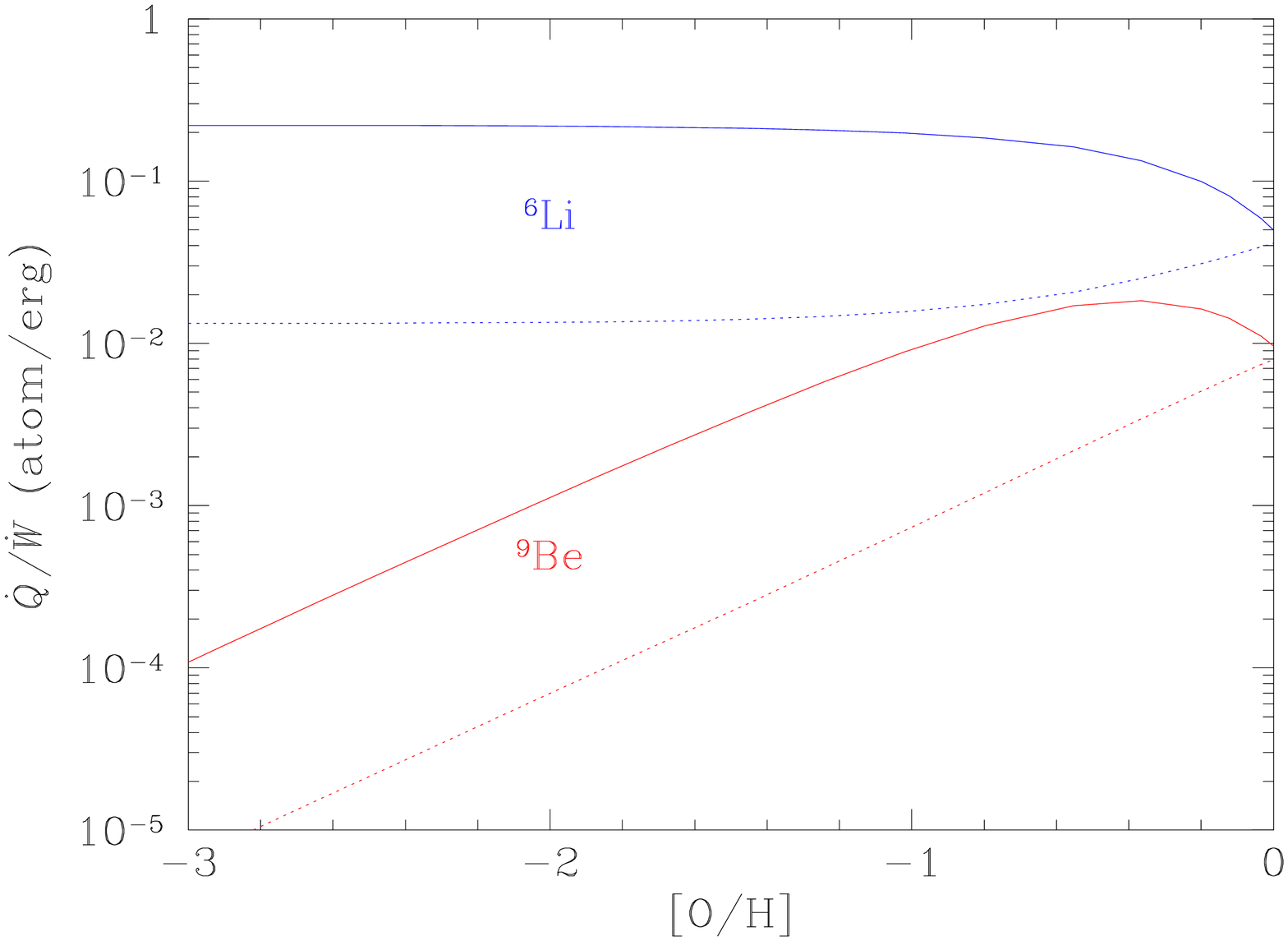,angle=270,width=0.9\textwidth} 

A plot of $\dot{Q}_{\rm Be}/\decr$ for our model. 
{\em Dotted lines}: constant $N_{\rm H}$; 
{\em full lines}: evolving $N_{\rm H}$.

\item
\label{fig:Be/O}
\psfig{file=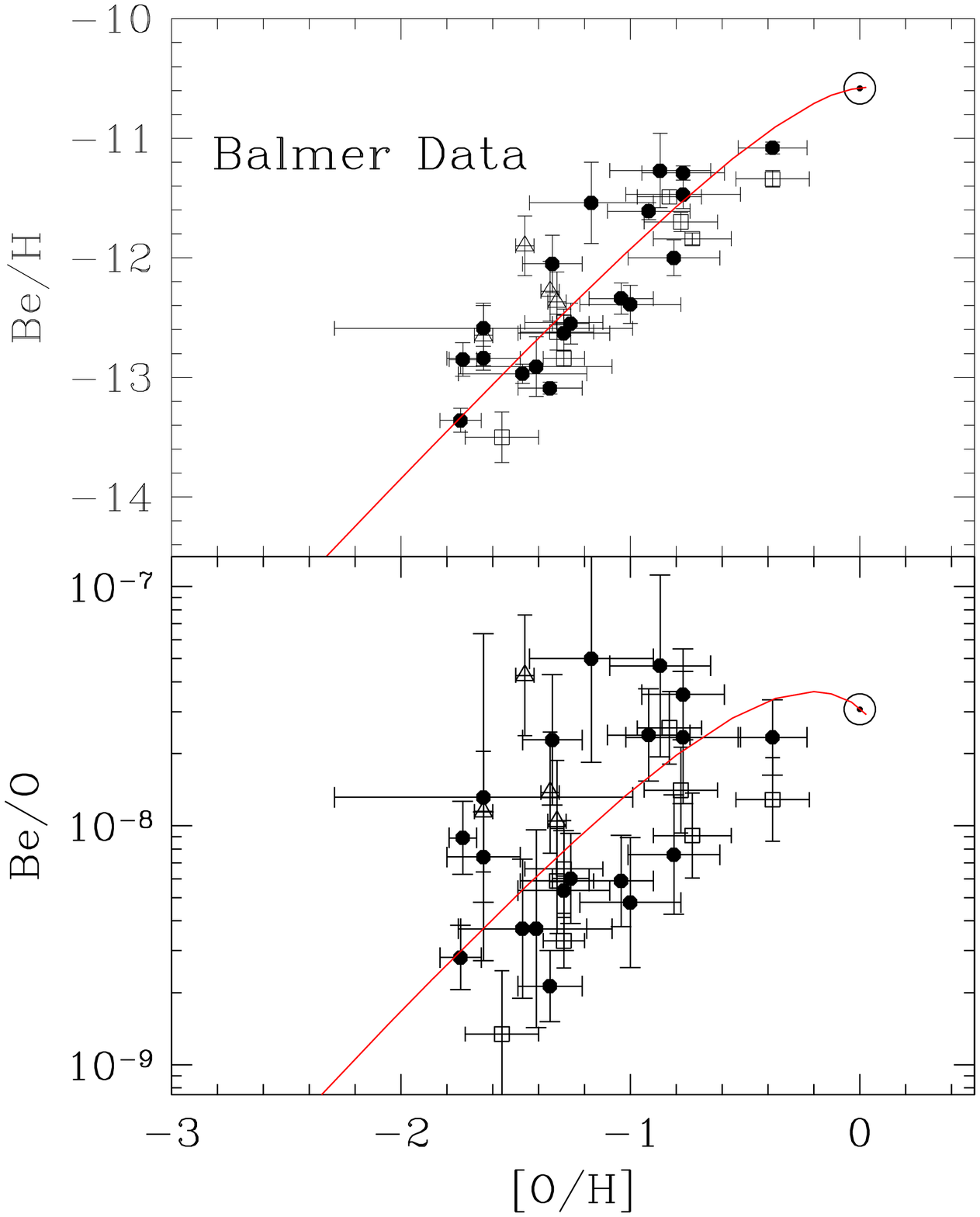,angle=0,height=0.8\textheight} 

Be and O data, derived from a consistent set of stellar
atmospheres using the ``Balmer'' method described in 
Fields \etal\ \pcite{fovc}.  The curve is for a model
in which Be is produced by
standard Galactic cosmic rays.

\item
\label{fig:energetics}
\psfig{file=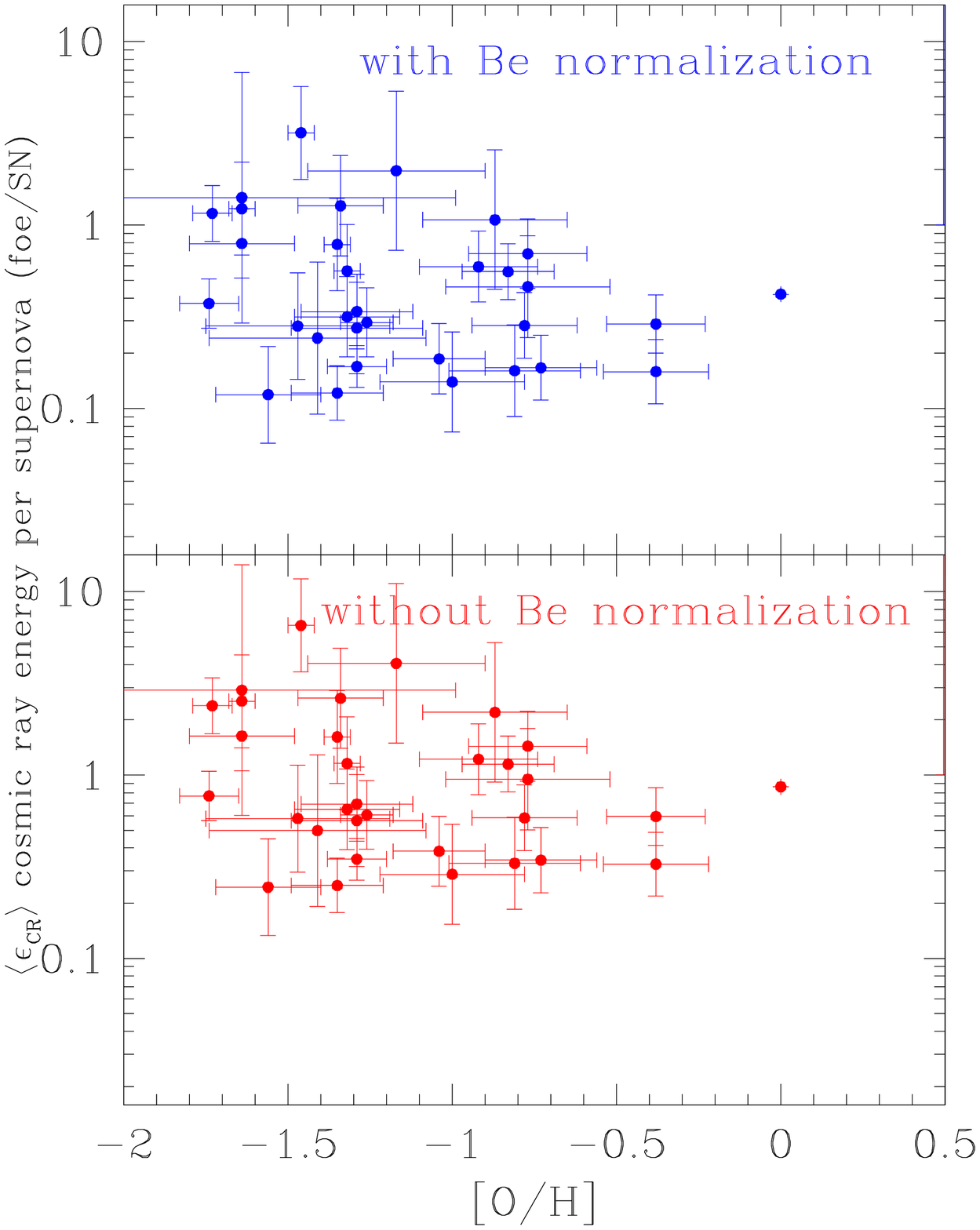,angle=0,height=0.8\textheight} 
Energetic requirements inferred point-by-point from the Be-O
data and the model shown in Figure \pref{fig:Be/O}.
The upper panel is for the standard model which normalizes the
Be point to solar; the lower panel does not use this
scaling.

\end{enumerate}


\begin{thebibliography}{bbdgp}

\bibitem[1997]{bv} Berezhko, E.\ 
G.\ and V{\"o}lk, H.\ J.\ 1997, Astroparticle Physics, 7, 183 

\bibitem[1993]{bk} Boesgaard A. \& King J.R. 1993, AJ, 106, 2309

\bibitem[1990]{bbdgp}
Brerezinski\u{\i}, V.S., Bulanov, S.V., Dogiel, V.A.,
Ginzburg, V.L., \& Ptuskin, V.S. 1990, Astrophysics of Cosmic Rays,
(Amsterdam:  North-Holland)

\bibitem[1994]{RICH} Buckley, J., Dwyer, 
J., Mueller, D., Swordy, S.\ and Tang, K.\ K.\ 1994, \apj, 429, 736 

\bibitem[1995] {cac} Cass\'e, M., Lehoucq, R., \& Vangioni-Flam, E. 1995,
Nature, 373, 38

\bibitem[1998]{connell}
Connell, J.\ J.\ 1998, \apjl, 501, L59 

\bibitem[1998]{crb} Combi, J.A., Romero, G.E., \& Benaglia, P.
1998, A\&A, 333, L91

\bibitem[1999]{dbb}Dragicevich, P.\ M., Blair, D.\ G., \& 
Burman, R.\ R.\ 1999, \mnras, 302, 693 

\bibitem[1989]{dmv} Drury, 
L.\ O., Markiewicz, W.\ J.\ and Voelk, H.\ J.\ 1989, \aap, 225, 179 

\bibitem[1990]{efsgj} Engelmann, J.\ J., 
Ferrando, P., Soutoul, A., Goret, P.\ and Juliusson, E.\ 1990, \aap, 233, 
96 

\bibitem[1999a]{fo99} Fields, B.D., \& Olive, K.A. 1999a, \apj, 516, 797

\bibitem[1999b]{fo99b} Fields, B.D., \& Olive, K.A. 1999b, New Astron., 4, 255

\bibitem[2000]{fovc} Fields, B.D., \& Olive, K.A., 
Vangioni-Flam, E. \& Cass\'e, M.  2000, \apj, 540, 930 (FOVC)

\bibitem[1987]{moises} Garcia-Mu\~{n}oz, M., 
Simpson, J.A., Guzik, T.G., Wefel, J.P., \& Margolis, S.H. 
1987, \apjs, 64, 269

\bibitem[1982]{hjk} Henderson, A.P., Jackson, P.D., \& Kerr, F.J.
1982, ApJ, 263, 116

\bibitem[1998]{hlr} Higdon, B., Lingenfelter, R.E., \& Ramaty, R. 1998,
ApJ, 509, L33 

\bibitem[1998]{igr} Israelian, G., Garc\'{\i}a-L\'{o}pez, R.J.,
\& Rebolo, R.  1998, ApJ, 507, 805

\bibitem[1982]{janni} Janni, J.F. 1982, Atomic Data
Nucl.\ Data Tables, 27, 34

\bibitem[1995]{koy} Koyama, K., \etal\ 1995, Nature, 378, 255

\bibitem[1994]{lfmw}
Lukasiak, A., Ferrando, P., McDonald, F.\ B.\ 
and Webber, W.\ R.\ 1994, \apj, 423, 426 

\bibitem[1970]{ns} Northcliffe, L.C., \& 
Schilling, R.F., Nucl.\ Data Tables, A7, 233

\bibitem[1997]{mori} Mori, M. 1997, ApJ, 4478, 225

\bibitem[2000]{par} Parizot, E. 2000, \aap, in press (astro-ph/0006099)

\bibitem[1997]{rklr} Ramaty, R., Kozlovsky, B., Lingenfelter, R.E.,
\& Reeves, H. 1997, ApJ, 488, 730

\bibitem[1998]{rkl} Ramaty, R., Kozlovsky, b., \& Lingenfelter, R. E.
1998, Physics  Today, 51, no. 4, 30

\bibitem[1999]{rl} Ramaty, R., \& Lingenfelter, R.E. 1999, in
LiBeB Cosmic Rays and Related X- and Gamma-Rays, eds.\ Ramaty
\etal, ASP, Vol. 171, p. 104

\bibitem[2000]{rslk} Ramaty, R., Scully, S.T., Lingenfelter, R.E., \& Kozlovsky, B.
2000, ApJ, 534, 747 

\bibitem[1970]{rfh} Reeves, H., Fowler, W.A., \& Hoyle, F. 
1970, Nature, 

\bibitem[1970]{rrga} 
Ryter, C., Reeves, H., Gradsztajn, E., \& Audouze, J.
1970, A\&A, 8, 389

\bibitem[1986]{scalo} Scalo, J.\ M.\ 1986, 
Fundamentals of Cosmic Physics, 11, 1 

\bibitem[1988]{sgm}
Simpson, J.\ A.\ \& Garcia-Munoz, M.\ 1988, Space Sci.\ Rev., 46, 205 

\bibitem[2000]{tak} Takeda, Y., Takada-Hidai, M., Sato, S., Sargent,
W.L.W., Lu, L., Barlow, T.A., Jugaku, J. 2000, astro-ph/0007007

\bibitem[1998]{tanimori} Tanimori, T., \etal\ 1998, ApJ, 497, L25

\bibitem[1997]{tdh} Timmes, 
F.\ X., Diehl, R.\ \& Hartmann, D.\ H.\ 1997, \apj, 479, 760 

\bibitem[2000]{vca}
Vangioni-Flam, E., Cass\'{e}, M., \& Audouze, J.
2000, Phys.\ Reports, 333-334, 365

\bibitem[1998] {vcfo} Vangioni-Flam, E., Cass\'e, M.,
Fields, B.D., \&  Olive, K.A. 1996, ApJ, 468, 199

\bibitem[1998] {vroc} Vangioni-Flam, E., Ramaty, R., Olive, K.A., \&
Cass\'e, M. 1998, 
 AA 337, 714

\bibitem[1998]{webber} Webber, W.R., 1998 \apj, 506, 329

\end{thebibliography}
\end{document}